\documentclass[useAMS,usenatbib]{mn2e}
\usepackage{amssymb}

\usepackage{Times}

\usepackage{graphicx,color}
\usepackage[fleqn]{amsmath}
\usepackage{bbm}
\def\d{{\rm d}}
\def\ylm{{Y_l^m}}
\def\msun{M_{\odot}}
\def\x{{\bmath x}} 
\def\v{{\bmath v}} 
\def\w{{\bmath w}}
\def\b#1{\bmath#1} 
\def\E{{\cal E}}   
\def\p{\partial}
\def\Var{{\rm Var}} 
\def\expect#1{\left\langle #1\right\rangle}
\def\vari#1{\Var\expect{#1}}
\def\I{{\mathbbmss{1}}}

\def\st{\sin \theta}

\title[Multi-mass schemes for $N$-body simulations]
{Multi-mass schemes for collisionless $N$-body simulations}
 \author[M. Zhang and S. J. Magorrian]
  {Mimi Zhang \thanks{E-mail: zhang@thphys.ox.ac.uk}
  and 
  John Magorrian  \thanks{E-mail: magog@thphys.ox.ac.uk}
  \\ 
  Rudolf Peierls Centre for Theoretical Physics, 1 Keble Road, Oxford
  OX1 3NP}
\begin{document}

\date{}

\pagerange{\pageref{firstpage}--\pageref{lastpage}} \pubyear{2007}

\maketitle
 
\label{firstpage}

\begin{abstract}
  We present a general scheme for constructing Monte Carlo
  realizations of equilibrium, collisionless galaxy models with known
  distribution function (DF) $f_0$.  Our method uses importance
  sampling to find the sampling DF $f_s$ that minimizes the
  mean-square formal errors in a given set of projections of the DF
  $f_0$.  The result is a multi-mass $N$-body realization of the
  galaxy model in which ``interesting'' regions of phase-space are
  densely populated by lots of low-mass particles, increasing the
  effective $N$ there, and less interesting regions by fewer,
  higher-mass particles.

  As a simple application, we consider the case of minimizing the shot
  noise in estimates of the acceleration field for an $N$-body model
  of a spherical Hernquist model.  Models constructed using our scheme
  easily yield a factor $\sim100$ reduction in the variance in the
  central acceleration field when compared to a traditional
  equal-mass model with the same number of particles.  When evolving
  both models with a real $N$-body code, the diffusion coefficients in
  our model are reduced by a similar factor.  Therefore, for certain
  types of problems, our scheme is a practical method for reducing the
  two-body relaxation effects, thereby bringing the $N$-body
  simulations closer to the collisionless ideal.
\end{abstract}


\section[]{Introduction}
\label{sec:intro}



There are two types of $N$-body simulations in stellar dynamics.
In {\em collisional} simulations each of the $N$ particles represents
an individual star.  This type of simulation is most often used to
model the evolution of star clusters in which discreteness effects,
such as two-body relaxation, are important.

When it comes to modelling galaxies, however, the number of stars is
large enough and the dynamical time is long enough that these
discreteness effects are usually unimportant.
In the limit of a very large number $N$ of bodies, stars and dark matter
particles move in a smooth mean-field potential $\Phi(\b x;t)$ and
behave as a collisionless fluid in six-dimensional phase-space
\citep[][BT87]{BT:1987}, the (mass) density at any point $(\b x,\b v)$ being 
given by the distribution function (hereafter DF) $f(\b x, \b v; t)$.  
The time-evolution of the DF is described by the Collisionless
Boltzmann Equation (hereafter CBE).  Therefore, in a {\em collisionless}
$N$-body simulation the $N$ particles do not correspond to real stars;
instead they provide a Monte Carlo realization of the smooth
underlying DF, from which one can estimate the potential $\Phi(\b x,t)$.
By integrating the orbits of these particles, one is solving the CBE
by the method of characteristics 
(\citealt{HO:1992}, HO92; \citealt{LCB:1993}, LCB1993)

In reality, no simulation is perfectly collisionless because Poisson
noise in the estimates of $\Phi(\b x, t)$ inevitably leads to
numerical diffusion in particles' orbits.  To reduce this noise, it
is important to make $N$, the number of particles in the simulation, as
large as possible.  Unfortunately, the cost of running an $N$-body
code scales at least linearly with~$N$, so increasing $N$ also makes
the simulation more expensive to run.  The good news is that
alternative, more sophisticated weapons are available for use in the
fight against small $N$ limitations.  A collisionless $N$-body code is
essentially a Monte Carlo method and so should be amenable to
well-known variance-reduction methods such as importance sampling
\citep[e.g.,][]{Press:1992}.

In this paper we present a generally-applicable, essentially
model-independent method for constructing $N$-body realizations of
isolated model galaxies in equilibrium, suitable for use as initial
conditions (hereafter ICs) in collisionless simulations.  Our scheme
uses importance sampling to find a sampling DF $f_s$ that minimizes
the
mean-square uncertainty in a chosen set of projections of the DF
$f_0$.  For example if modelling bar evolution, one might be most
interested in following the detailed evolution of the DF around the
strongest resonances.  It is natural then to try to increase the
sampling density near these regions by populating them with
lots of low-mass particles.  Outside these interesting regions,
however, one must
also have enough particles to maintain accurate
estimations of the force field which governs the evolution of the
system as a whole.

The paper is organized as follows.
After reviewing the basics of Monte Carlo integration and the
connection between CBE and $N$-body simulations, we explain our
multi-mass formulation in section~\ref{sec:choosefs}.
With the notable exception of some heuristic multi-mass schemes (e.g., 
\citet{SHQ:1995}, hereafter SHQ95, \citet{WK:2007}, \citet{Sellwood:2008} and
\citet{zemp:2007}), most other IC-generation schemes have used equal-mass
particles.
In section~\ref{sec:implementation} we give an example of using our
scheme to suppress fluctuations in the monopole component of
acceleration in a spherical galaxy model.  We calculate formal
estimates of the noise in $N$-body models constructed the equal-mass
scheme (section~\ref{sec:fsequal}) and SHQ95's method
(section~\ref{sec:fssig}) and compare them to our our own scheme in
section~\ref{sec:fsmulti}.  In section~\ref{sec:nbodyrealizations} we
test
how well our realizations behave in practice when evolved using a real
$N$-body code.  Finally, section~\ref{sec:conclusions} contains a
summary of the prerequisites for our scheme, along with possible
scientific applications.

\section[]{Formulation}
\label{sec:basics}

\subsection{Monte Carlo integration}

For later reference we recall some of the basic ideas
\citep[e.g.,][]{Press:1992} in using Monte Carlo methods to evaluate
integrals, such as
\begin{equation}
  \label{eq:simplemcintegral}
  I = \int_D f\,\d V,
\end{equation}
of a known function~$f$ over a domain~$D$.  We first consider the case
where $D$ has unit volume: $\int_D\,\d V=1$.  Then, given $N$ points,
$x_1\ldots x_N$, drawn uniformly from~$D$ we can estimate
\begin{equation}
 I\simeq \frac1N\sum_{i=1}^N f(x_i)\equiv \expect{f}.
\end{equation}
The variance in this estimate
\begin{equation}
  {1\over N^2}\sum_{i=1}^N\left[f(x_i)-\expect{f}\right]^2\to
  \frac1N\left\{\int_D f^2\,\d V - \left[\int_Df\,\d V\right]^2\right\}
\end{equation}
as $N\to\infty$.

Now let us relax the assumption that $D$ has unit volume and, instead
of drawing points uniformly from~$D$, let us take $N$ points drawn
from a sampling distribution $f_s$.  We assume that $f_s$ is
normalised: $\int_D f_s\,\d V=1$.  Making a straightforward change of
variables and using the result above, it follows that the
integral~(\ref{eq:simplemcintegral}) can be estimated as
\begin{equation}
\label{eq:mcmean}
  I \simeq {1\over N} \sum_{i=1}^N\frac{f(x_i)}{f_s(x_i)},
\end{equation}
and that the variance in this estimate is approximately 
\begin{equation}
\label{eq:mcvar}
  \Var I = \frac1N\left[\int_D \frac{f^2}{f_s}\,\d V -
    I^2\right].
\end{equation}

\subsection{$N$-body simulations and the CBE}

The following description of the connection between $N$-body
simulations and the CBE borrows heavily from LCB93.  We assume that
the galaxy has total mass~$M_*=1$ and that the mass density of stars
in phase-space is given by a DF~$f(\b x,\b v;t)=f(\b w;t)$, where $\b
w\equiv (\b x,\b v)$, normalised so that
\begin{equation}
  \label{eq:DFnorm}
  \int f(\b w)\,\d ^6\b w=1.
\end{equation}

The evolution of the DF is governed by the CBE,
\begin{equation}
   \label{eq:CBE}      
  \frac{\p f}{\p t} + \b v\cdot\frac{\p f}{\p\b x} -
  \frac{\p\Phi}{\p\b x}\cdot\frac{\p f}{\p\b v}=0.
\end{equation}
It conserves phase-space density, so that
\begin{equation}
  \label{eq:liouville}
  f(\b w(t);t)=f(\b w_0;0),
\end{equation}
where $\b w(t)$ is the path traced by an individual particle, with $\b
w_0\equiv \b w(t=0)$.  As HO92 and LCB93 point out, in a collisionless
$N$-body simulation one is solving the CBE for these $\b w(t)$ by
integrating the characteristic equation,
\begin{equation}
  \label{eq:charac}
  \frac{\d t}{1} = \frac{\d\b x}{\b v} = \frac{\d\b v}{\b a},
\end{equation}
and using Monte Carlo integration to estimate the acceleration
\begin{equation}
  \label{eq:accelsx}
  \b a(\b x;t) \equiv -{\p\Phi\over\p\b x}=
-G\nabla\int {f(\b w';t)\over |\b x-\b x'|}\,\d^6\b w'.
\end{equation}
From~(\ref{eq:mcmean}) it
follows that 
\begin{equation}
  \label{eq:accels}
  \b a(\b x;t) \simeq -G\nabla\sum_{i=1}^N \frac{m_i}{|\b x-\b x_i|},
\end{equation}
corresponding to a distribution of $N$ point particles with masses
\begin{equation}
  \label{eq:masses}
  m_i = \frac 1N \frac{f(\b w_i;t)}{f_s(\b w_i;t)}.
\end{equation}
These $m_i$ clearly depend on the choice of sampling DF
$f_s$.  The simplest choice is $f_s(\b w;t)=f(\b w;t)=f(\b
w_0;0)$, in which case all particles have equal masses $m_i=1/N$.
However, one is free to
tailor the choice of $f_s$ to suit the particular problem under study.

The singularities in~(\ref{eq:accels}) at $\b x=\b x_i$ yield
estimates of $\b a(\b x)$ that suffer from unacceptably large scatter;
in fact, they correspond to the direct accelerations appropriate for a
collisional $N$-body code!  So, in practice collisionless simulations
do not use (\ref{eq:accels}) directly, but instead obtain $\b a(\b x)$
using techniques (e.g., softened force kernels, grid methods or
truncated basis-function expansions) that reduce the scatter by
removing the singularities.
More generally, eq.~(\ref{eq:accels}) provides only the most
simple-minded estimate of the integral~(\ref{eq:accelsx}), and one has
some leeway in how one reconstructs $f(\b w;t)$ from the discrete
realization furnished by the $N$ particles.  Of course, the
reliability of any sensible reconstruction will be wholly dependent on
how well the DF is sampled.

\subsection{Observables}
\label{sec:windowfn}

What constitutes a ``good'' choice of sampling density~$f_s$?  The DF
$f$ is a high-dimensional probability density and itself is not
measurable.  We are usually only interested in coarse-grained
projections of the DF,
\begin{equation}
  \label{eq:mom}
  \expect{Q_i} \equiv \int f(\b w) Q_i(\b w)\,\d^6\b w,
\end{equation}
where the kernels $Q_i(\b w)$ are some functions of the phase-space
co-ordinates $(\b x, \b v)$.  For the purposes of the present paper,
we consider a ``good'' sampling scheme to be one that minimizes the
uncertainty in the estimates of some given set of $\expect{Q_i}$.
Apart from some general guidance, we do not address the question of
how best to choose these $Q_i$, which usually requires some experience
of the particular problem at hand.

We now give some examples.  It is helpful to introduce the indicator function 
\begin{equation}
    \label{eq:indicator}
    \I_V(\w)\equiv
    \begin{cases}
      1, & \mbox{if $\w\in V$} \cr 
      0. & \mbox{otherwise.}
    \end{cases}
\end{equation}
Then a particularly simple but important choice of kernel is
\begin{equation}
\label{eq:simpleq}
  Q_i(\b w)= \I_{V_i}(\b w),
\end{equation}
for which $\expect{Q_i}$ measures the mass inside a volume $V_i$.  For
many problems one might choose some of the $V_i$ to surround important
resonances in phase-space, so that $\expect{Q_i}$ measures the
phase-space density around the resonances.
With appropriate choices of projection kernel $Q_i$, the
expression~(\ref{eq:mom}) includes quantities such as the galaxy's
density profile, its velocity moments or even its projected
line-of-sight velocity distributions.

More fundamentally, an $N$-body model should provide a good estimate
of the galaxy's acceleration field.  Therefore we recommend that many
of the $\expect{Q_i}$ be used to measure at least the monopole
component of the galaxy's acceleration field at a range of points.
This can be achieved using (\ref{eq:simpleq}) with spherical volumes
$V_i$ centred on $\b x=0$ for a range of radii $r_i$, encompassing
all velocity space for $|\b x|<r_i$.  Similarly, one can include
higher-order multipole components of the galaxy's acceleration field
by choosing a slightly more complicated projection kernel $Q_i$ (see
equation~\ref{eq:qilm} below).

\subsection{Optimal sampling scheme}
\label{sec:choosefs}

The problem we address in this paper is the following.  We wish to
construct an equilibrium $N$-body realization of a galaxy model with
some known DF $f_0$.  Specifically, we seek ICs that
faithfully represent some projections,
\begin{equation}
  \label{eq:proj}
  \expect{Q_i} = \int f_0Q_i\,\d^6\b w,
\end{equation}
of this DF, for a set of $n_Q$ kernels $Q_i(\b w)$.  What is the
``best'' choice of sampling DF $f_s$ given this $f_0$ and choice of
kernels~$Q_i$?

More formally, from~(\ref{eq:mcvar}) the uncertainty in a Monte Carlo
estimate of $\expect{Q_i}$ obtained using $N$~particles drawn from the
sampling distribution~$f_s$ is given by
\begin{equation}
  \label{eq:deltaq}
  \vari{Q_i} = \frac1N\left[\int \frac{f_0^2}{f_s}Q_i^2\,\d^6\b w -
    \expect{Q_i}^2\right].
\end{equation}
Notice that, unlike most introductory textbook examples of Monte Carlo
methods, we have $n_Q$ such estimates but just one~$f_s$.  We seek a
normalised sampling DF $f_s$ that minimizes the mean-square
fractional uncertainty
\begin{equation}
  \label{eq:S}
  S \equiv \sum_{i=1}^{n_Q} (\delta Q_i)^2
\end{equation}
where $\delta Q_i$, the formal relative uncertainty in a measurement of
$Q_i$, is given by
\begin{equation}
  \label{eq:deltaqi}
  (\delta Q_i)^2 \equiv \frac{ \vari{Q_i} }{ \expect{Q_i}^2 }.
\end{equation}
Of course there are many other possible measures of the
``goodness'' of some choice of~$f_s$.

One can immediately use the Euler--Lagrange equation to show that
choosing 
\begin{equation}
  \label{eq:eulersoln}
  f_s^2(\b w) \propto f_0^2(\b w)\sum_{i=1}^{n_Q}\frac{Q_i^2(\b w)}{\expect{Q_i}^2}
\end{equation}
extremizes~(\ref{eq:S}), the proportionality constant being set
by the constraint that $f_s$ should be normalized, $\int f_s\,\d^6\b
w=1$.
This direct solution is flawed, however, since for most interesting
choices of $Q_i$ the resulting $f_s$ depends on orbit phase; using
this $f_s$ the masses of particles sampling a given orbit would vary
along the orbit!  Therefore in practice we use a slightly less direct
approach.

We partition phase space into $n_f$ cells and write
$\tau_j$~\footnote{Note that we use $V$ to denote subvolumes of
  phase-space be used in the calculation of the
  projections~(\ref{eq:mom}) of the DF $f_0$, and $\tau$ for the
  subvolumes used in the discretization of the sampling DF $f_s$.}
for the phase-space volume enclosed by the $j^{\rm th}$ cell.  We
parametrize $f_s$ as
\begin{equation}
  \label{eq:discretefs}
  f_s(\b w) = \sum_{j=1}^{n_f} \frac{\I_{\tau_j}}{a_j}f_0(\b w),
\end{equation}
so that within the $j^{\rm th}$ phase-space cell $f_s$ is given by
$f_0(\b w)/a_j$.  For the equilibrium models considered, it is natural to
choose $\tau_j$ to be cells in integral space.  Substituting this
$f_s$ into~(\ref{eq:deltaqi}) yields
\begin{equation}
  \label{eq:deltaqidiscrete}
  (\delta Q_i)^2 = \frac1N\left[\sum_{j=1}^{n_f}a_jH_{ij}-1\right],
\end{equation}
where
\begin{equation}
  \label{eq:Hij}
  H_{ij}\equiv \frac{\int_{\tau_j} f_0Q_i^2\,\d^6\b w}{\expect{Q_i}^2}.
\end{equation}
If we further define 
\begin{equation}
  \label{eq:H}
  H_j\equiv \sum_{i=1}^{n_Q}H_{ij},
\end{equation}
then the mean-square fractional uncertainty~(\ref{eq:S}) becomes
\begin{equation}
  \label{eq:Sdiscrete}
  S = \frac1N\left[\sum_{j=1}^{n_f}a_jH_j-n_Q\right].
\end{equation}
Our goal is to find the coefficients $a_j$ that minimize this $S$,
subject to the constraint that the resulting $f_s$ be normalised.  The
normalisation constraint is that
\begin{equation}
  \label{eq:normaj}
  \int f_s\,\d^6\b w = \sum_{j=1}^{n_f}\frac{I_j}{a_j}=1,
\end{equation}
where
\begin{equation}
  \label{eq:Ij}
  I_j\equiv \int_{\tau_j}f_0\,\d^6\b w.
\end{equation}
Using the method of Lagrange multipliers, the coefficients of the
``best'' sampling DF obtained by minimizing~(\ref{eq:Sdiscrete})
subject to the constraint~(\ref{eq:normaj}) are simply
\begin{equation}
  \label{eq:aj}
  a_j = \sqrt\frac{I_j}{H_j}\,\sum_{k=1}^{n_f}\sqrt{I_kH_k},
\end{equation}
which is just the direct solution~(\ref{eq:eulersoln}) in disguise,
but averaged over the phase-space cells $\tau_j$ and correctly
normalized.
This averaging means that the resulting $f_s$ will be smooth, provided
that none of the kernels $Q_i$ pick out specific regions of integral
space.

Substituting the $f_s$ given by (\ref{eq:discretefs}) into
(\ref{eq:masses}), we have that particles in phase-space cell $\tau_j$
have masses $m_j=a_j/N$.  One can therefore easily impose additional,
direct constraints on the masses of particles within a subset of the
phase-space cells~$\tau_j$; simply repeat the minimisation of
(\ref{eq:Sdiscrete}) subject to (\ref{eq:normaj}) while holding the
relevant subset of the $a_j$ fixed at the desired values.  For
example, when generating an $N$-body realization of a dark-matter halo
model inside which one intends to embed a disk of light particles, one
might want to ensure that those halo particles passing through the
disk have the same mass as the disk particles.
Of course, a more pedestrian approach would be to introduce additional
kernels $Q_i$ to pick out the relevant parts of integral space.  We
caution, however, that we have not tested how well such a ``bumpy''
$f_s$ would work in practice; the tests we present later all involve
smoothly varying sampling distributions.

\subsection{ICs for $N$-body model}
\label{sec:ics}

Together with $f_0$, the coefficients $a_j$ completely determine the
sampling DF $f_s$ of the form~(\ref{eq:discretefs}). In particular, it
reduces to the conventional equal-mass case when all $a_j=1$.

We apply the following sequence of steps $N$ times to draw particles
from this $f_s$, thereby constructing an $N$-body realization of the
galaxy model:
\begin{enumerate}
  \renewcommand{\theenumi}{(\arabic{enumi})}
\item Choose one of the $n_f$ cells at random, the probability of
  choosing the $j^{\rm th}$ cell being given by $I_j/a_j$.  Let $i$ be
  the index of the chosen cell.
\item Assign a mass $m_i\equiv f_0(\b w_i)/Nf_s(\b w_i)=a_i/N$ to the
  particle.
\item Within the $i^{\rm th}$ cell, draw $\b x_i$ from its density
  distribution, $\rho_i=\int f_0\I_{\tau _i}\d^6\b w$.  For the
  special case of a spherical galaxy, one can precompute the
  cumulative mass distribution $M_i(r)$ for each of the $n_f$ cells
  and use this to draw a radius $r_i$, followed by angles $\theta_i$
  and $\phi_i$.
\item Use an acceptance-rejection method to draw $\b v_i$ from $f_0(\b
  x_i,\b v)$ at this fixed value of $\b x_i$.
\end{enumerate}

\section[]{An example}
\label{sec:implementation}
In this section we use a simple galaxy model to demonstrate our scheme.
Our galaxy model is spherical and isotropic, with density profile \citep{H:1990}
\begin{equation}
  \label{eq:hernquist}
\rho(r) = \frac {M_*}{2\pi r(r+a)^3},
\end{equation}
total mass $M_*$ and scale radius $a$.  By Jean's theorem, the model's
DF $f(\b x, \b v)$ depends on $({\x,\v})$ only through the binding
energy per unit mass $\E$.  \citet{H:1990} gives an expression for
$f(\E)$.

We want to construct an $N$-body realization of this model that
minimizes the mean-square error in the monopole component of the
acceleration averaged over many decades in radius, from $r_{\rm min} =
10^{-4} a$ up to $r_{\rm max} = 10^2 a$. To achieve this we choose
kernels $Q_i=\I_{V_i}(r)$ that measure the mass enclosed within a
sequence of 25 spheres centred on the origin, with radii $r_i$ spaced
logarithmically between $r_{\rm min}$ and $r_{\rm max}$.  We use
(\ref{eq:deltaqidiscrete}) to calculate the formal uncertainty $\delta
M_i$ in the enclosed mass for a range of discretized sampling
densities of the form~(\ref{eq:discretefs}), including~(\ref{eq:aj}).

To implement this, we first of all partition integral space $(\E,J^2)$
onto a regular $n_f = n_\E \times n_{X}$ grid.  The $n_\E$ energies
$\E_j$ are chosen to match the potential $\E_j=\Psi(r_j)$ with $r_j$
logarithmically spaced between $10^{-6}a$ and $10^3a$.  At each
$\E_j$, there are $n_X$ values of $X_{jk}$ running linearly from $0$
to $1$, where $X_{jk} = J_k(\E_j)/J_c(\E_j)$ is the orbital angular
momentum normalised by the circular angular momentum at energy $\E_j$
\footnote{${X}=J(\E)/J_c(\E)$ is a measure of an orbit's circularity;
  orbits with $X=1$ are perfectly circular, while those with $X=0$ are
  perfectly radial.}.  These choices ensure that our $f_s$ samples
well the interesting parts of phase-space. For the calculations below
we take $n_\E\times n_X=200\times100$, although a coarser grid
(e.g., $50\times25$) would suffice.  Having defined our projection
kernels $Q_i=\I_{V_i}$, we use (\ref{eq:proj}) to calculate the expected
values of enclosed mass $\expect{M_i}$ and the ancillary quantities
$H_{ij}$ (from eq.~\ref{eq:Hij}),  and use these to obtain the formal
uncertainties $\delta M_i$ in~(\ref{eq:deltaqidiscrete}).

\begin{figure}
  \begin{center}
    \includegraphics[width=0.48\textwidth] {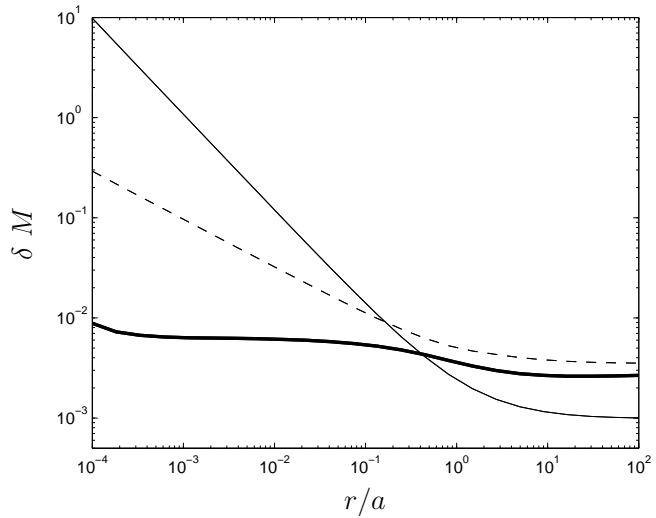}
  \end{center}
  \caption{Formal relative errors $\delta M\equiv (\vari M
    )^{1/2}/\expect{M}$ (eq.\ \ref{eq:deltaqi}) in the monopole
    component of the potential of a Hernquist model (eq.\
    \ref{eq:hernquist}) constructed using the same number $N=10^6$ of
    particles, but drawn from different sampling DFs.  The heavy solid curve
    plots results for our tailored sampling DF.  For comparison, we
    also show results for the conventional equal-mass scheme (light
    solid curve) and Sigurdsson et al.'s (1995) heuristic multi-mass
    scheme (dashed curve).}
  \label{fig:err}
\end{figure}

\subsection{Comparison with other schemes: formal errors}
\label{sec:comparison}
Before applying our method, we study two other schemes: the
conventional equal-mass scheme and the multi-mass scheme of \citet{SHQ:1995}.

\subsubsection{The conventional equal-mass scheme}
\label{sec:fsequal}
The most common (albeit implicit) choice of sampling density is
$f_s=f_0$, which corresponds to setting all $a_j=1$ in our
equation~(\ref{eq:discretefs}).  All particles then have the same
$1/N$ mass.  For our example Hernquist model the fraction of particles
within radius~$r$ is $r^2/(a+r)^2$, so that less than $1\%$ of the
particles are within $0.1 a$. 
As shown in figure~\ref{fig:err}, for this $f_s$ the formal
uncertainty $\delta M(r)$ rises steeply towards the centre.  Although
this scheme produces accurate
estimates of the galaxy's potential outside the scale radius $a$, it
performs poorly in the interesting $r^{-1}$ central density cusp.

\subsubsection{Sigurdsson et al.'s multi-mass scheme}
\label{sec:fssig}
\citet{SHQ:1995} have used an interesting heuristic scheme to improve
the resolution of $N$-body models near galaxy centres.  In effect,
they use an anisotropic sampling function of the form
(\ref{eq:discretefs}) with coefficients
\begin{equation}
  a(\tau) \equiv B\times
  \begin{cases}
    \left(\frac{r_{\rm peri} (\tau) }a\right)^{\lambda} & \hbox{if
      $r_{\rm peri}<a$},\cr 1 & \hbox{otherwise},
  \end{cases}
\end{equation}
where $r_{\rm peri}(\tau)$ is the smallest pericentre radius of any
orbit from the phase-space cell~$\tau$, and the constant $B$ is chosen
to normalize~$f_s$.  When the parameter $\lambda=0$, then $a_j = 1$
and the sampling DF $f_s$ is identical to $f_0$.  Increasing $\lambda$
improves the sampling of the cusp by increasing the number density of
particles having pericentres $r_{\rm peri}<a$. Consequently, as
$r\to0$ the number density of particles rises more rapidly than the
mass density, permitting better resolution in the centre.  To balance
this increase in number density, each particle is assigned a mass
$f_0/Nf_s= a_j /N$ so that the phase-space mass density is still given
by the desired~$f_0$.

The dashed curve in figure~\ref{fig:err} shows the formal error
$\delta M(r)$ in our implementation of their scheme for $\lambda=1$.
Their scheme does much better than the conventional equal-mass scheme
at small radii $r\ll a$, at the cost of a slightly noisier monopole at
$r\gtrsim a$.

\begin{figure}
  \begin{center}
    \includegraphics[width=0.48\textwidth] {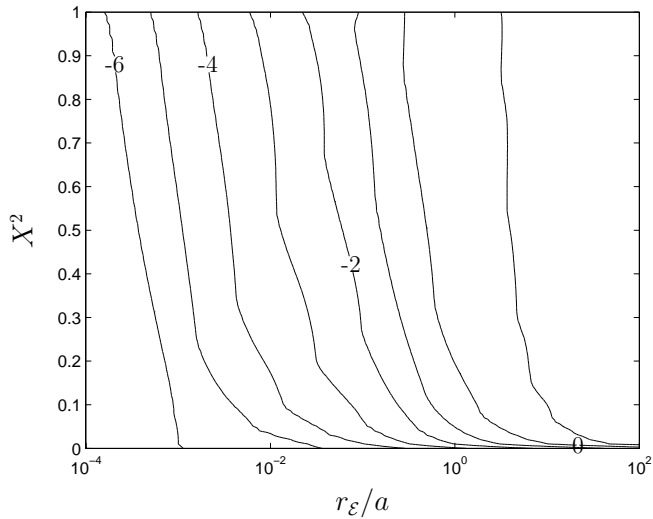}%
  \end{center}
\caption{Contour map of $\log_{10} (f_s/f_0)$ for the optimal
  muli-mass sampling scheme of section~\ref{sec:fsmulti}; notice the
  strong enhancement in low angular momentum and large energy (small
  $r_{\E}$) region. These correspond to orbits with small peri-centre
  radii.}
\label{fig:ajk}
\end{figure}

\subsubsection{Our scheme}
\label{sec:fsmulti}
It is encouraging to see that SHQ95's multi-mass scheme does, to some
extent, improve mass resolution at small radius. However, as shown in
figure~\ref{fig:err}, $\delta M$ at $r=10^{-4}a$ is still almost two
orders of magnitude larger than at $r=a$.  Can we achieve even better
results by carefully designing an $f_s$ that generates a flat $\delta
M(r)$ across a large range of radii?

The $f_s$ given by our optimal choice of coefficients ~(\ref{eq:aj})
is plotted in figure~\ref{fig:ajk}.  It is qualitatively similar to
SHQ95's results, in the fact that it samples densely the low-angular
momentum parts of phase-space.  The detailed shape of the function is
different, however, and the thick solid curve in figure~\ref{fig:err}
shows that our scheme provides much better estimates of the monopole
components of the acceleration at small radii; in fact, the formal
error $\delta M(r)$ varies by only a factor $\sim 4$ over six decades
in radius.

\begin{figure}
  \begin{center}
    \includegraphics[width=0.48\textwidth]{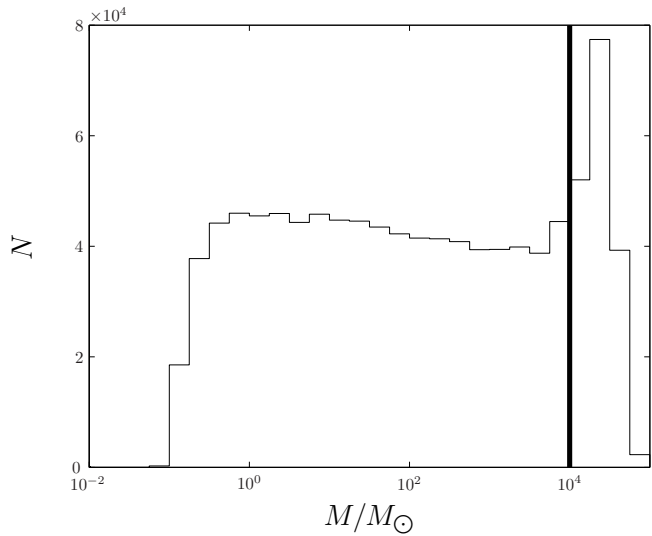}
  \end{center}
  \caption{Histogram of particle masses from an $N =
    10^6$ multi-mass realization of a Hernquist galaxy, scaled to a
    total mass $M_\star=10^{10}\msun$.  The span of 8 decades in mass
    gives sub-solar mass resolution in ``interesting'' regions of
    phase space.  In contrast, in an equal-mass realization all
    particles would have mass $10^4\msun$ (thick solid line).}
  \label{fig:nptle}
\end{figure}

\begin{figure}
  \begin{center}
    \includegraphics[width=0.48\textwidth]{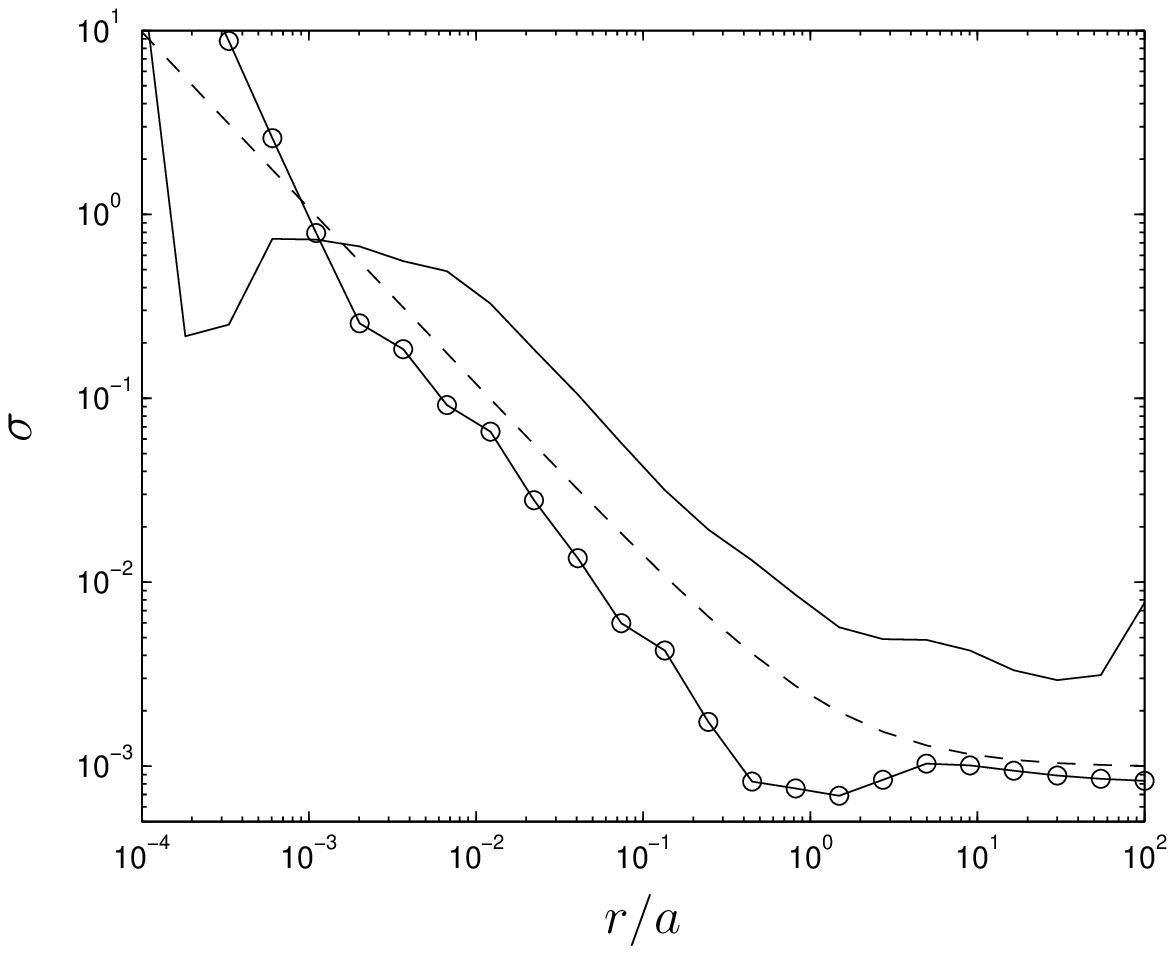}
    \\
    \vskip1cm
    \includegraphics[width=0.48\textwidth]{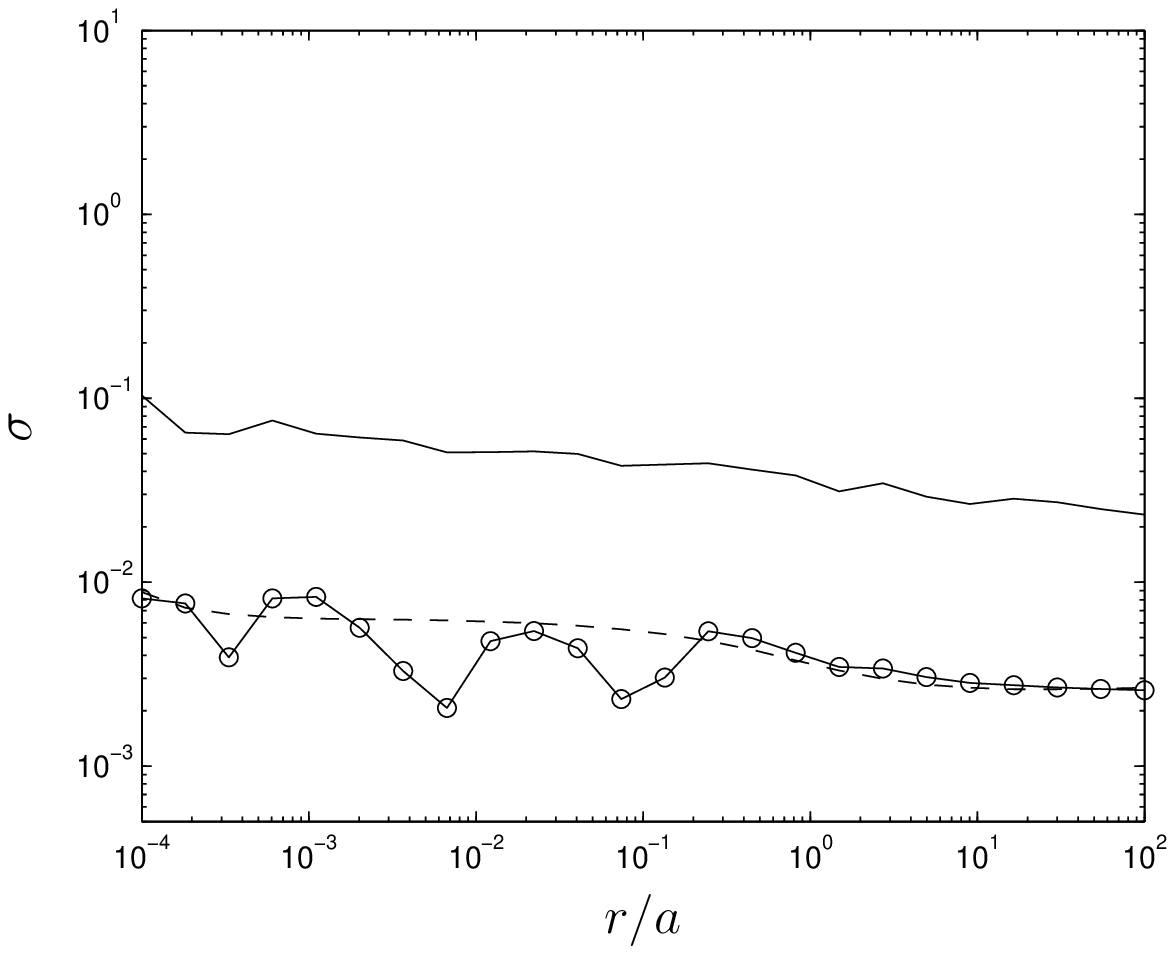}
  \end{center}
  \caption{ RMS fractional deviations in acceleration (solid line), in
    mass (circle-solid line) together with its analytical value
    $\delta M$ in dashed line; top panel for equal-mass model and
    bottom panel for multi-mass model.  For this single acceleration
    calculation, we include $16$ levels of refinements and therefore
    all the values should be believable outside $\epsilon_{\rm min}
    \approx 10^{-4}$.  }
  \label{fig:delta}
\end{figure}

\subsection{$N$-body realizations}
\label{sec:nbodyrealizations}
Figure~\ref{fig:nptle} shows the spectrum of masses obtained using the
algorithm detailed in \S\ref{sec:ics} to draw $N=10^6$ particles from this
optimal $f_s$.  Unlike the conventional scheme which would give all
particles the same $10^4 \msun$ mass if assuming $M_{*} = 10^{10}
\msun$, our multi-mass scheme assigns a range of masses between
$10^{-2}\msun$ and $10^{6}\msun$ (8 decades), with many low-mass
particles in the central region.

As a simple sanity check of our formal estimates of the errors in the
monopole, we count the mass of particles within the same spheres $V_i$
used to calculate $\delta M_i$. The deviations from the mass profile
of the target Hernquist model are consistent (figure~\ref{fig:delta})
with the expected values of $\delta M$ from
equation~(\ref{eq:deltaqidiscrete}).

Ultimately, the purpose of our sampling scheme is to improve the
numerical modelling of collisionless galaxies close to equilibrium
using full $N$-body integrations.
To test how well our scheme succeeds at this task, we use the
particle-multiple-mesh code Grommet \citep{Magorrian:2007} to compare
the evolution of our multi-mass models against equal-mass ones.
Below, we adopt $N$-body units $G=M=a=1$.  

\subsubsection{How well is the acceleration field reproduced?}
\label{sec:test1} 

An important unsolved problem is how best to estimate the
accelerations~(\ref{eq:accelsx}) given a discrete $N$-particle
realization of the underlying DF~$f$.  The most sophisticated
approaches to this problem (e.g., \citet{D:2001} and references
therein) have focused on finding softening kernels that minimize the
errors in the acceleration field given a static distribution of $N$
{\em equal-mass} particles.
In the present paper we do not investigate how different softening
lengths or softening kernels affect our multi-mass models.  We simply
adopt a nested series of boxes with boundaries at $|\b
x|=100\times2^{-i}$ with $i=0,\ldots,20$, each box covered by a
$60^3$ mesh.  As one moves to smaller length scales the effective
softening length decreases, with $\epsilon_{\rm min} = 200/60\times
2^{-20} \approx 10^{-4} $.

Figure~\ref{fig:delta} shows the  fractional error in the radial
component of the acceleration field returned by Grommet, in addition
to the fractional error in enclosed mass.  There is an approximately
constant offset between these two quantities for equal- and multi-mass
realization. Since our ICs here have been tailored to
minimize the variance in the monopole component of the acceleration
field, how important is our neglect of the higher-order
multipoles?

In terms of multipole moments, the radial component of the
acceleration is (e.g., BT87)
\begin{align}
  \label{eq:arlm}
  a_r(r,\theta,\phi) & = 4 \pi G \sum_{lm} \frac{\ylm}{2l+1}
  \times   \nonumber \\
  &\left[ - \frac{l+1}{r^{l+2}} \int_0^r \rho_{lm}(r') r'^{l+2} \d r'
    + l r^{l-1} \int_r^\infty \rho_{lm}(r') \frac{\d r'} {r'^{l-1}}
  \right]
\end{align}
where
\begin{equation}
  \rho_{lm}(r) =  \int_0^{2 \pi} \!\!\d \phi
  \int_0^{\pi} \!\! \d \theta \st \,
  \ylm^*(\theta,\phi) \rho(r,\theta,\phi).
\end{equation}
This can be rewritten as
\begin{equation}
  a_r(r, \theta, \phi) =  - \frac{4 \pi G}{r^2} \sum_{lm} 
  \expect{M_{lm}(r)}\ylm(\theta, \phi),
\end{equation}
where $\expect{M_{lm}(r)}$ are given by
\begin{equation}
\expect{M_{lm}(r)} = \int f_0(\b w')M_{lm}(r,\b w')\,\d^6\b w'
\end{equation}
with projection kernels
\begin{equation}
\label{eq:qilm}
    M_{lm}(r,\b w') =  \ylm^*(\theta',\phi') \left[
         \frac{l+1}{2l+1} \frac{r'^l}{r^l}  \I_{V(r)}(\b w')
       - \frac{l}{2l+1} \frac{r^{l+1}}{r'^{l+1}}
       \I_{\overline{V(r)}}(\b w')
   \right],
\end{equation}
where $V(r)$ encompasses all phase-space points having radii
less than the (real-space) radius~$r$, and $\overline {V(r)}$ is
its complement.  For our spherical galaxy,
\begin{equation}
  \expect{M_{lm}(r)} =  
  \int  f_0(w') \, M_{lm}(r,\b w') \,\d^6 \b w'
  =
  \begin{cases}
    M(r), & \hbox{if $l=m=0$}\cr
    0, & \hbox{otherwise}.
  \end{cases}
\end{equation}
The corresponding variance in $a_r(r)$ for an $N$-body realization
drawn from some choice of $f_s$ is
\begin{equation}
  \label{eq:vararlm}
  \vari{a_r(r)} =  \frac{4 \pi G}{r^2} \sum_{lm} 
                      \vari{M_{lm}(r)} \ylm(\theta, \phi),
\end{equation}
where, from~(\ref{eq:deltaq}),
\begin{align}
  \vari{M_{lm}(r)} = \frac1N\Big[ &
  \int \frac{f_0^2(\b w')}{f_s(\b w')} M_{lm}(r,\b w') ^2 \d^6 \b w'\\
  &\quad - \expect{M_{lm}(r)}^2\Big].
\end{align}

Similarly, the variance in the tangential component of acceleration
field can be achieved by using projection kernels
\begin{equation}
  \label{eq:varatlm}
  \vari{a_{\theta, \phi}(r)} =  \frac{4 \pi G}{r^2} \sum_{lm} 
  \vari{M_{lm}^t(r)} \left[\ylm(\theta, \phi)\right]^{'}_{\theta,\phi},
\end{equation}
where
\begin{equation}
  \label{eq:qilmt}
    M_{lm}^t(r,\b w') =  \ylm^*(\theta',\phi') \left[
         \frac{1}{2l+1} \frac{r'^l}{r^l}  \I_{V(r)}(\b w')
       + \frac{1}{2l+1} \frac{r^{l+1}}{r'^{l+1}}
       \I_{\overline{V(r)}}(\b w')
   \right].
\end{equation}

So, given any choice of $f_s$, we can use the expressions above to
calculate the contribution of the higher-order multipole moments to
the formal errors in the acceleration.  We find that, as we progressively include
more terms, our estimate of the formal
$\vari{a_r(r)}$ approaches the actual errors observed in the $N$-body
realization.

Alternatively, we can find our optimal sampling DF~$f_s$ by minimizing
\begin{equation}
S \equiv \sum_{i=1}^{n_Q} 
\frac{ \displaystyle{\sum_{l=0}^{l_{\rm max}}\sum_{m=-l}^l \vari{M^i_{lm}}} }{\expect{M_i}^2},
\end{equation}
truncated at say $l_{\rm max}=2$.  Notice that this new $S$ reduces to
the old one in eq.~(\ref{eq:S}) when $l_{\rm max}=0$, but otherwise
includes additional terms with $l>0$, each weighted by the monopole
component $l=0$.  On increasing $l_{\rm max}$ from 0 to 2, the formal
$\vari{a_r(r)}$ increases but the shape of the curve remains
approximately unchanged and there are no noticeable differences in the
resulting $f_s$.  Therefore, our neglect of higher-order multipole
moments is justified, at least in the present case, provided one bears
in mind that the errors in the full acceleration field are going to be
larger by an approximately constant factor than what one would
estimate from the monopole component alone.

\subsubsection{How well are integrals of motion conserved?}
\label{sec:test2} 
This paragraph describes the details of a full $N$-body
implementation. Using both equal and multi-mass schemes, we draw
$10^6$ particles with radii between $10^{-3} < r <10^2$. In order to
suppress slightly deviance from symmetry (the odd terms of
higher-order multipoles) and remove any intrinsic transient in linear
momentum \citep[see also][]{MD:2005}, ICs $(\x,\v)$ are extended to
include the mirror distribution by reflecting each of the $10^6$
particles with $(\b x,\b v)\to(-\b x,-\b v)$.  The full ICs then have
$N=2\times10^6$ particles.  Taking the efficiency of integration into
consideration, only a $12$-level nested series of boxes each covered
by a $60^3$ mesh is used, together with a single time-step of
$2\times10^{-4}$.
Therefore, we expect our numerical results to be trustworthy at radii
greater than a few times $10^{-3} $.
\begin{figure}
  \begin{center}
    \includegraphics[width=0.48\textwidth]{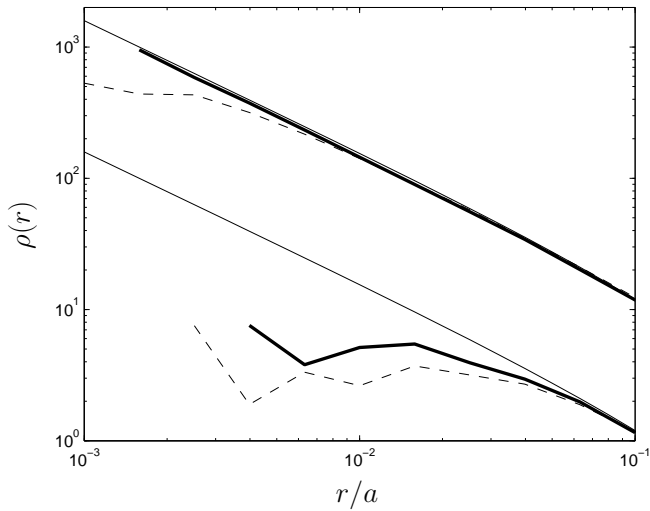}
  \end{center}
  \caption{Inner density profile of a multi-mass (top set of curves)
    and an equal-mass (lower, offset by 10 vertically) realizations of
    the same Hernquist model evolved for 200 time units using the
    Grommet $N$-body code.  The ICs in each case are plotted as the
    heavy solid curves.  The dashed curves show profiles after the
    model has been been evolved for 200 time units.  }
  \label{fig:testprof}
\end{figure}

Figure~\ref{fig:testprof} plots the inner density profiles of both
realizations after evolving each for $200$ time-units (or 300 circular
orbit periods at $r=0.01$).  The lack of particles at small radius $r
\sim 10^{-2}$ in the equal-mass realization means that the initial
model is out of exactly-detailed equilibrium and causes the central
density profiles to flatten.  In contrast, the density profile in the
multi-mass case is always much better behaved there.

It is interesting to examine what is going on at the level of
individual orbits.  Both realizations begin with spherical symmetry
and remain spherical, apart from the effects of Poisson noise.  The
amount of diffusion in the angular momentum~$J$ of each particle's
orbit serves as a strong gauge of relaxation effects.  This is
complicated by the fact that many particles in isotropic models being
considered here have $J(t=0) \simeq 0$. In such cases, even a small
change in $J(t>0)$ would yield a large fractional change when measured
in respect to its initial value. To circumvent this artificial
problem, for each particle we measure the change in angular momenta
relative to its circular value at $t=0$ using
\begin{equation}
   \label{eq:deltaX2}
   \Delta X^2_i  =  \left[ \frac{J_i(t)-J_i(0)}{J_c(\E_i)} \right]^2 \frac1t.
\end{equation}
Binning particles by energy and calculating the mean $\Delta X^2_i$
within each energy bin gives us the time-averaged diffusion rate $\delta X^2(\E)$.
\begin{figure}
  \centerline{
    \includegraphics[width=0.48\textwidth]{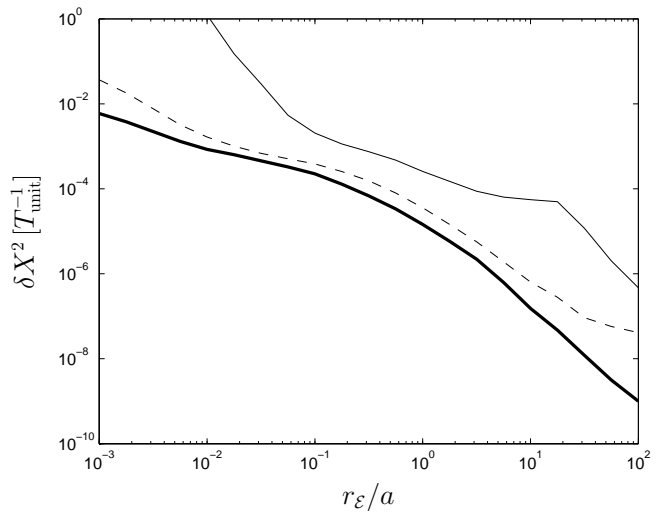}
  }
  \caption{Time-average diffusion rate $\delta {X^2}$
    (eq.~\ref{eq:deltaX2}) against energy labelled radius $r_{\E}$
    measured between $t=0$ and $t=200$ for a multi-mass realization
    evolved using Grommet (thick curve), falcON (dashed) and for an
    equal-mass realization evolved using Grommet (thin curve).
  }
  \label{fig:diffX2a}
\end{figure}
As shown in figure~\ref{fig:diffX2a}, both models suffer diffusion,
but due to the enhancement of particle numbers and hence the
smoothness of potential field in the central region,
diffusion in the multi-mass scheme is suppressed by two orders of
magnitude across the whole system.

As a further test of the robustness of our multi-mass scheme, we have
evolved our multi-mass ICs using the tree code FALCON \citet{D:2000}
with a single interparticle softening radius of $10^{-3}$, comparable
to the finest mesh size used in the Grommet runs.  The dashed curve in
figure~\ref{fig:diffX2a} plots the resulting $\delta X^2$; our scheme
works just as well for tree codes as it does for mesh codes, although
the variable softening in Grommet does slightly decrease the amount of
diffusion.  This is not surprising, since the only difference between
the two runs is the approximations used to estimate the
accelerations~(\ref{eq:accelsx}).

\begin{figure}
  \centerline{
    \resizebox{86mm}{!}{\includegraphics{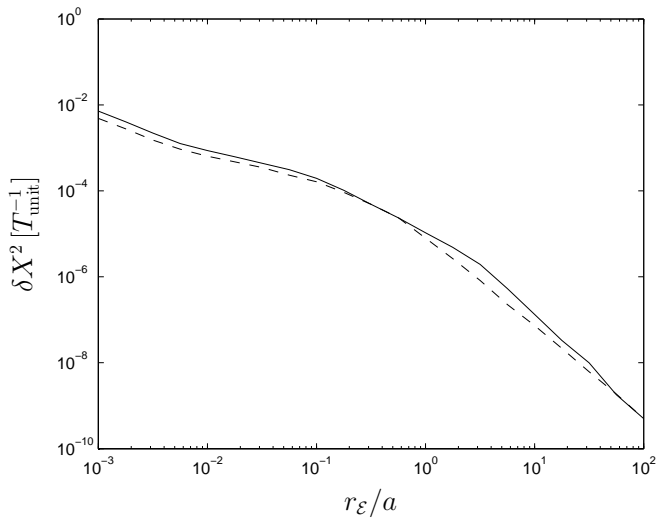}} }
  \caption{$\delta {X^2}$ in multi-mass simulations for
    particles with $X^2 < 0.1$ (dashed curve) and $X^2 > 0.9$ (solid). }
    \label{fig:diffX2b}
\end{figure}

In any model with a broad spectrum of particle masses, a natural
question is what happens if heavy bodies from the outskirts visit the
centre full of light mass elements and vice versa.
To address this issue, we have measured the $\delta X^2(\E)$ of
equation~(\ref{eq:deltaX2}) but, instead of considering all particles
of a given $\E$, we compare the diffusion of particles on
radially-biased orbits with $X^2 < 0.1$ to those on nearly-circular
orbits with $X^2>0.9$.  As shown in figure~\ref{fig:diffX2b}, there
are no systematic differences between them.  The reason for this is
simply that particles with $X\simeq0$ spend most of their time at
apocentre, the apocentre radius being only a factor $\sim2$ larger
than the radius of a circular orbit of the same energy.  Nevertheless,
a particle with $X\simeq0$ will affect all of the more tightly bound
orbits as it plunges through the centre of the galaxy, but our
measured diffusion rates account for this. 

\section[]{Conclusions}
\label{sec:conclusions}

We have presented a general multi-mass scheme to construct Monte Carlo
realizations of collisionless galaxy models with known steady-state
DFs $f_0$. 
The scheme uses importance sampling to find the tailored sampling DF
$f_s$ that minimizes the sum of mean-square uncertainties in a given
set of observable quantities of the form (\ref{eq:proj}).  Although our
method works for any reasonably general collisionless $N$-body code,
we note that there are three conditions that must be satisfied before
it can be applied successfully:
\begin{enumerate}
\item The system should be in a steady state, or close to one.
\item The DF $f_0$ should be quick and cheap to evaluate, either
  numerically or analytically.  The calculation of $f_s$ is not much
  more demanding for axisymmetric or triaxial galaxies than for
  spherical isotropic models.  Finding $f_0$ for such systems is,
  however, nontrivial since one rarely has sufficient knowledge of the
  underlying potential's integrals of motion, but suitable flattened
  DFs do exist, including the standard axisymmetric two-integral
  $f(\E,L_z)$ models and also rotating triaxial models such as those
  used in, e.g., \citet{ber:2006}.  An alternative way of constructing
  flattened multi-mass realizations would be to apply
  \citet{holley:2002}'s adiabatic sculpting scheme to a spherical
  $N$-body model constructed using our scheme.
\item Finally, the utility of our multi-mass scheme depends critically
  on the selection of the projection kernels $Q_i(\w)$.
\end{enumerate}
Point (ii) is a well-known and longstanding problem, but the final
condition is new.  It is probably best addressed by experimenting with
different sets of kernels, especially since it is easy to test the
consequences of modifying them.  Nevertheless, there are cases in
which modest physical insight offers some guidance on choosing the
$Q_i$.  Here are some examples.

{\bf Galaxies with central massive black holes}\qquad It is
straightforward to extend our treatment of self-consistent galaxy
models to models containing a central black hole (hereafter MBH). By
choosing kernels (as in section~\ref{sec:implementation}) to measure
{\em the monopole component of galaxy's force field} and choosing
$f_s$ to minimize their mean-square fractional uncertainty, one also
achieves better spatial and mass resolution within the sphere of
influence of the black hole.

{\bf Loss-cone problems}\qquad The rate of supplying stars into MBH's
loss-cone is an important ingredient in galaxy models with central
MBHs. A thorough understanding of collisionless loss-cone refilling
mechanisms and accurate estimates of the resulting refilling rates are
particularly critical for the prediction of astrophysical quantities
such as the timescale of binary MBH merger
\citep{BBR:1980,Y:2002,MM:2003}, the tidal disruption rate of stars
\citep{SU:1999,MT:1999,WM:2004}. When using $N$-body simulations to
study such loss-cone problems, one is often interested in {\em stars
  on low angular momentum orbits} and can therefore choose kernels to
pick out such loss-cone phase-space for detailed modelling, while
simultaneously maintaining accurate estimates of the galaxy's
acceleration field.

{\bf Sinking satellites}\qquad \citet{KM:2004} demonstrate the
significance of using equilibrium $N$-body realizations of satellite
models when investigating the effect of tidal stripping of CDM
substructure halos (satellites) orbiting inside a more massive host
potential.  Besides the shape of the background potential and the
amount of tidal heating, the mass-loss history is very sensitive to
the detailed density profile of the satellite itself.  One can
therefore make one step further from equal-mass realizations by
designing kernels to pick out {\em orbits that pass through the tidal
  radius}, while again maintaining an accurate estimate of the
satellite's acceleration field.

\section*{Acknowledgements}
We thank James Binney for helpful discussions and an anonymous referee
for comments that helped improve the clarity of this paper.  MZ's work
is supported by Dorothy Hodgkin Postgraduate PPARC-BP Awards.  JM
thanks the Royal Society for financial support.

\bsp

\label{lastpage}

\end{document}